# Energy-Efficient Routing Algorithm for Wireless Sensor Networks: A Multi-Agent Reinforcement Learning Approach

P. Soltani, M. Eskandarpour, A. Ahmadizad, H. Soleimani

Iran University of Science & Technology, Department of Electrical Engineering

**Abstract—** Efficient energy management is essential in Wireless Sensor Networks (WSNs) to extend network lifetime and ensure reliable data transmission. This paper presents a novel method using reinforcement learning-based cluster-head selection and a hybrid multi-hop routing algorithm, which leverages Q-learning within a multi-agent system to dynamically adapt transmission paths based on the energy distribution across sensor nodes. Each sensor node is modeled as an autonomous agent that observes local state parameters—such as residual energy, distance to sink, hop count, and hotspot proximity—and selects routing actions that maximize long-term energy efficiency. After computing the optimal paths, each sensor aggregates sensed data and forwards it through intermediate nodes to a selected transmitter node, chosen based on the highest remaining State of Charge (SoC), thereby avoiding premature node depletion. To promote efficient learning, a carefully designed reward function incentivizes balanced load distribution, hotspot avoidance, and energy-aware forwarding while maintaining signal quality. The learning process occurs either in a decentralized manner or via a cloud-based controller that offloads computation in large-scale deployments. Moreover, the RL-driven routing decisions are fused with classical graph-based methods—Minimum Energy Routing Algorithm (MERA) and Minimum Spanning Tree (MST)—to optimize energy consumption and load balancing. Simulations confirm that the proposed approach significantly improves node survival rate, reduces SoC variance, and enhances network resilience, making it a scalable and adaptive solution for energy-constrained WSNs in dynamic sensor deployments and IoT applications.

## I. INTRODUCTION

Wireless Sensor Networks (WSNs) play a critical role in modern applications such as precision agriculture [1], healthcare monitoring [2], industrial automation [3], and smart city infrastructure. [4] These networks consist of spatially distributed sensor nodes equipped with microcontrollers, transceivers, and limited battery power. [5] Given their deployment in remote or hard-to-reach areas, recharging or replacing node batteries is often infeasible. [6] Consequently, energy efficiency is a primary design concern, as it directly impacts network longevity and reliability in data delivery. [7]

To address energy constraints in WSNs, various routing strategies have been proposed. [8] Clustering-based protocols like LEACH and its extensions aim to reduce transmission distance by aggregating data at designated cluster-heads. [9] Hierarchical methods such as LEACH-T enable multi-hop data forwarding across clusters. [10] Bio-inspired algorithms, including Ant Colony Optimization and Bee Colony Optimization, leverage decentralized control to adapt to dynamic network states. [11] Reinforcement learning (RL) techniques have also emerged to optimize routing policies based on feedback from the network environment. [12] Other approaches, such as mobile charging, load-balancing schemes, and data freshness metrics like Age of Information (AoI), aim to improve network performance through more diverse perspectives. [13]

Despite these advancements, existing methods often suffer from critical limitations [14]. Clustering protocols, while reducing transmission cost, can lead to uneven energy consumption due to the heavy burden on cluster-heads, requiring frequent re-clustering. [15] Static and heuristic-based methods fail to adapt to changes in traffic, energy distribution, or node failures. Even some RL-based solutions focus on short-term optimization without adequately addressing long-term energy balance or overall network resilience. [16] Moreover, few approaches leverage global knowledge or predictive strategies to guide adaptive routing decisions. [17]

To overcome these challenges, this paper proposes a hybrid, energy-aware routing framework based on Q-learning and enhanced with graph-theoretic strategies. [18] Each sensor node operates as an independent learning agent, perceiving local network conditions via a comprehensive state vector— including normalized residual energy (SoC), sink distance, hop count, queue length, and more. [19] Two classical algorithms— Minimum Energy Routing Algorithm (MERA) and Minimum Spanning Tree (MST)—are integrated to improve robustness. MERA prioritizes routes through energy-rich nodes, while MST reduces congestion by minimizing total path cost. Their outcomes are fused using a weighted approach to dynamically adjust to network demands. The system supports centralized decision-making via a cloud server that collects periodic state vectors and distributes optimal routing policies. [20]

The main contributions of this work are as follows. First, we propose a novel, single-cluster reinforcement learning framework that enables dynamic multi-hop routing in WSNs, balancing local energy awareness with global route efficiency. [21] Second, we integrate the Minimum Energy Routing Algorithm (MERA) and Minimum Spanning Tree (MST) methods to create a robust hybrid routing strategy that both reduces energy concentration and avoids network bottlenecks. Third, we design a rich state representation for each node, capturing energy level, hop count, buffer usage, and neighborhood context to inform intelligent routing decisions. Fourth, a cloud-assisted mechanism is incorporated to offload computation from sensor nodes, enabling centralized policy learning and efficient role assignments. Finally, we validate the proposed framework through simulations in both 2D and 3D network topologies, demonstrating significant improvements in energy efficiency, node survival rates, and overall routing performance compared to traditional clustering techniques.

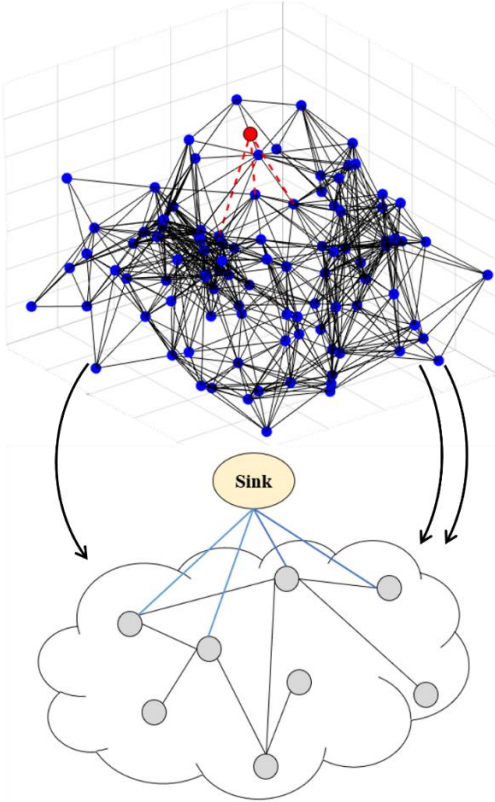

**Fig. 1.** Simulated wireless sensor network as a graph.

The rest of this paper is structured as follows. Section II provides a review of related work on WSN routing and reinforcement learning techniques. Section III introduces the system model and the formulation of the Q-learning framework. Section IV details the system model of the multi-agent reinforcement learning approach and in Section V the paper presents simulation results and performance comparisons with existing methods. Finally, Section VI concludes the paper and outlines directions for future research.

## II. RECENT RESEARCHES

Recent advancements in WSNs have highlighted the importance of energy-aware routing and adaptive decision-making. Early works by Frank [22] and Pritsker [23] approached the problem using multiple integrals to represent probabilistic quantities, which then required numerical methods for estimation. However, the computational complexity of these methods scaled rapidly. To address the growing computational burden, Mirchandani [24] introduced a reliability-based approach for evaluating link connectivity. This method, while efficient, assumes discrete random variables for edge lengths, which limits its general applicability.

To mitigate computational inefficiencies, several researchers turned to Monte Carlo simulation for probabilistic estimation. Notable contributions in this area include works by Fishman [25], Adlakha [26], and Sigal et al. [27]. Despite its flexibility, Monte Carlo simulation remains computationally intensive and demands prior knowledge of edge length distributions, which may not be available in practical settings.

In response to these limitations, alternative methods based on learning automata (LA) have been proposed. Beigy and Meybodi [28] introduced a distributed LA network for identifying stochastic shortest paths in packet-switched networks. This approach was later enhanced by Guo et al. [29], who introduced a hierarchical structure and refined convergence criteria to improve efficiency. Expanding on these ideas, Liu and Zhao [30] proposed a full-path feedback model incorporating a forced exploration strategy based on random barycentric spanners. Similarly, He et al. [31] employed a semi-bandit feedback model, which collects individual edge feedback rather than full-path observations, enabling more nuanced control over the trade-off between exploration and exploitation.

Other studies have applied algorithms from the multi-armed bandit (MAB) literature to related path optimization problems. For instance, the combinatorial upper confidence bound (CUCB) algorithm [32] and Thompson Sampling (TS) [33] have been adapted for scenarios where entire paths are treated as composite actions. CUCB constructs super-arms by selecting multiple basic actions and balancing uncertainty with expected rewards, while TS models the uncertainty of each action with Beta distributions and selects paths based on sampled estimates. In the context of wireless rechargeable sensor networks (WRSNs), Wei et al. [34] proposed the CSRL strategy to enhance wireless charging entity (WCE) path planning. Their hybrid approach combines simulated annealing for action selection with conventional reinforcement learning (RL) to determine optimal charging paths for all sensor nodes. However, traditional RL, which relies on a single agent, suffers from slow convergence in complex environments. To overcome this, Iima and Kuroe [35] introduced Swarm Reinforcement Learning (SRL), where multiple agents collaboratively learn by sharing Q-values. SRL significantly improves convergence rates and facilitates more efficient policy learning. Motivated by these advantages, the present study adopts SRL to enhance path planning performance in dynamic and distributed environments. In 2021, Chen et al. [36] proposed the Distributed Load-Balanced Routing (DLBR) approach, which distributes data packets based on both node energy levels and communication load, thus enhancing network durability. More recently, Zhang and Li (2022) [37] introduced the Adaptive Hierarchical Routing (AHR) algorithm, which dynamically adjusts the network's hierarchy structure in response to node energy, data generation rate, and location. AHR reduced routing overhead and improved scalability by up to 30% compared to traditional methods. Despite their advantages, these approaches remain limited in their adaptability to sudden changes in network state, and often rely on predefined structures or static decision-making. Our proposed method addresses these gaps by combining multi-agent reinforcement learning with global coordination and graph-based routing enhancements. Unlike DLBR and AHR, our framework dynamically adapts to network conditions using real-time local state vectors and a centralized Q-learning policy. In addition, our hybrid strategy incorporates MERA and MST to ensure energy-balanced routing and congestion avoidance, resulting in a more robust and efficient solution for long-term network performance.

Clustering has been widely used in WSNs to improve scalability and energy efficiency. Protocols such as LEACH introduced the idea of periodically electing cluster-heads to aggregate and forward data to the sink, significantly reducing the number of long-range transmissions. Variants like LEACH-T attempted to enhance routing by supporting limited multi-hop communication between cluster-heads and the base station. More recent studies have explored clustering in three-dimensional environments to better reflect real-world deployment scenarios. However, these methods suffer from several drawbacks: frequent re-clustering introduces high overhead, cluster-heads often experience rapid energy depletion, and energy usage becomes imbalanced across the network. Moreover, most clustering protocols rely on static or heuristic configurations that are not well-suited for dynamic WSN environments where node energy, data traffic, or topology may change over time.

To address the limitations of static routing, reinforcement learning (RL) has emerged as a promising solution for adaptive and data-driven decision-making in WSNs. Several works have proposed Q-learning-based routing mechanisms where nodes learn optimal paths based on feedback from their environment. These methods can dynamically respond to changing energy levels and network conditions, making them more robust than traditional heuristic approaches. However, most existing RL-based solutions focus on local optimization without considering long-term energy balance or global network efficiency. In addition, they often lack mechanisms for coordinating routing decisions across multiple nodes, leading to suboptimal energy usage and communication bottlenecks. In contrast, our proposed framework builds on multi-agent Q-learning with a centralized learning policy that combines local energy awareness with global route efficiency. Furthermore, by integrating MERA and MST algorithms into the routing process, we enhance decision robustness while maintaining scalability and adaptability.

### III. PROPOSED METHOD

In Wireless Sensor Networks (WSNs), sensing activity is often spatially non-uniform. Some areas generate significantly more data than others, causing certain nodes to become overburdened and deplete their energy quickly, while other nodes retain unused energy. This imbalance leads to connectivity loss and degraded network coverage over time. To address this issue, we propose an adaptive routing strategy based on multi-agent reinforcement learning (RL) that prolongs network lifetime by minimizing energy variance and avoiding premature node failures.

The core objective of our method is to achieve even energy utilization across all nodes. This is accomplished by using Q-learning at each node to make deliberate, energy-aware routing decisions. Rather than selecting routes purely based on shortest distance or minimum immediate energy cost, each node evaluates multiple paths and favors those that minimize the variance of State of Charge (SoC) among the participating nodes. This ensures that high-energy nodes are used more often for data forwarding, relieving overused nodes and balancing the overall load.

In the proposed single-cluster architecture, a single node—designated as the main transmitter—is dynamically selected to forward aggregated data to the sink. Figure 2 illustrates the complete workflow of the proposed energy-aware routing algorithm within a Wireless Sensor Network (WSN). Subfigure (a) presents the initial deployment of sensor nodes uniformly distributed across a two-dimensional field, where each blue dot denotes a sensor node. At this stage, communication links have not yet been established, and the figure solely represents the initial network topology. In subfigure (b), communication links are formed between nodes that are within mutual transmission range, thereby generating a connectivity graph. This graph defines the feasible set of routing paths and constitutes the action space for each node within the Q-learning framework. Subfigure (c) depicts the dynamic selection of the main transmitter node, identified in red, based on the maximum residual energy (State of Charge, SoC). This node is responsible for transmitting aggregated data to the sink. Concurrently, the green-highlighted region identifies a subset of source nodes that initiate data transmission within the network.

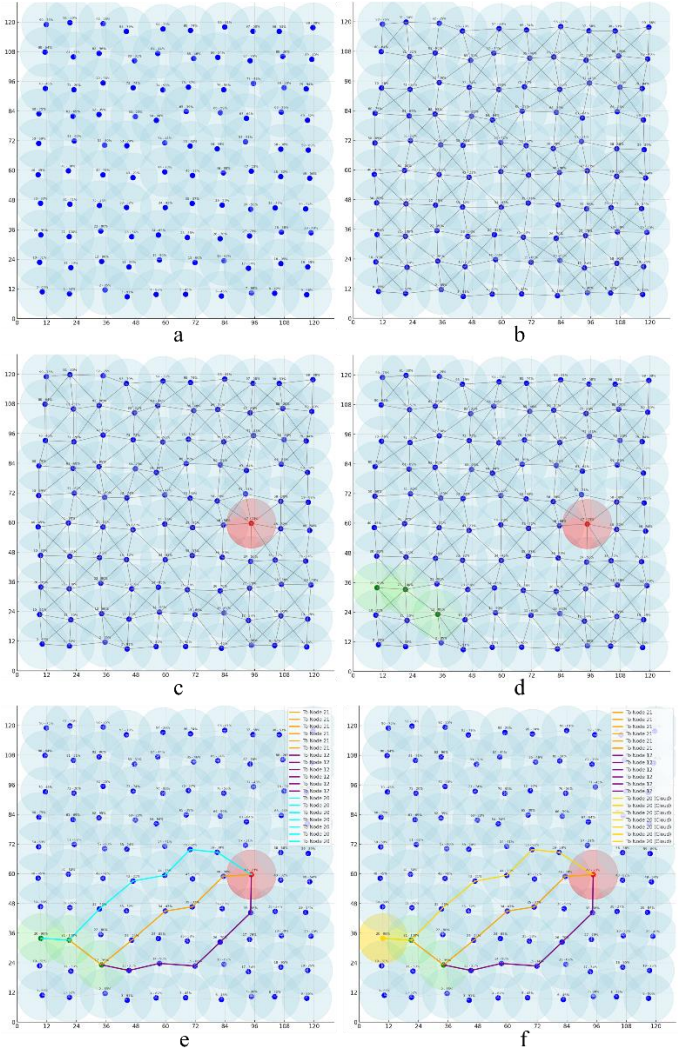

**Fig. 2.** Step-by-step visualization of the proposed energy-aware routing in a WSN. (a–b) Node deployment and communication graph. (c–d) Dynamic selection of the transmitter node (red) and active source nodes (green). (e–f) Energy-balanced multi-hop routing paths selected based on SoC variance and SNR constraints, with path colors indicating node energy levels.

Subfigure (d) represents the early phase of the routing process, where Q-learning begins to guide the routing decisions. The transmitter remains the same, while green source nodes begin evaluating paths to it using their learned Q-tables, taking into account both energy and connectivity. In subfigure (e), energy-aware multi-hop paths start to emerge. These paths are color-coded based on the SoC levels of the nodes they pass through. The algorithm prefers routes that include high-energy nodes and avoid low-energy ones, thereby minimizing SoC variance along each path. Finally, subfigure (f) shows the complete routing paths after full policy execution. These paths satisfy both the signal quality (SNR) and energy balance constraints. The diverse set of routes demonstrates the algorithm's adaptability, effectively distributing the network load and preventing the formation of energy hotspots.

Once the transmitter is selected, the remaining nodes must forward their data to it using multi-hop routing. Fig. 3 illustrates this pathfinding process. To select optimal paths, the algorithm simultaneously considers two key constraints:

1. Signal Quality Constraint – The total hop count must remain within a threshold to preserve end-to-end Signal-to-Noise Ratio (SNR).
2. Energy Balance Constraint – Among all valid paths, the algorithm selects the one that minimizes the standard deviation of SoC, distributing energy consumption more evenly.

This dual-objective routing strategy is implemented through Q-learning, where each node is modeled as an autonomous agent. At each decision point, a node observes its local state vector, which includes:

- normalized residual energy (SoC),
- distance to sink and transmitter,
- queue length (buffer usage),
- hop count estimate,
- hotspot proximity (how often the node was used in past routes),
- average energy of neighboring nodes,
- and the node's current role (sensor, forwarder, or transmitter).

Using this state, the node selects actions such as forwarding data, entering sleep mode, or requesting the transmitter role. The node's Q-table is updated using a reward function designed to promote long-term energy balance.

Specifically, the reward function provides higher values when:

- nodes with high energy are selected for routing,
- routes avoid energy hotspots,
- SoC variance is minimized,
- and data is successfully delivered with minimal intermediate hops.

To avoid overfitting and encourage exploration, we use an ε-greedy policy, where agents occasionally try less-optimal routes to gather new experiences. The Q-learning parameters—learning rate (α), discount factor (γ), and exploration rate (ε)—are empirically tuned for convergence and stability.

To reduce computation and communication overhead, especially in large-scale deployments, we apply a pre-filtering step to the network graph. This step removes physically infeasible links (e.g., those beyond communication range), thus shrinking the action space and accelerating learning. To formally express the energy-balancing objective, we define the SoC variance for a candidate path $p_i$ as:

$$(\sigma_1)^2 = \frac{1}{n_i - 1} \sum_{j \in p_i} (x_i - \bar{x})^2 \quad (1)$$

where $x_i$ is the SoC of node j along the path, and $\bar{x}$ is the average SoC of that path. Among all valid paths, the algorithm selects:

$$p^* = \underset{p_i}{\operatorname{argmin}}(\sigma_1)^2 \quad (2)$$

This ensures the algorithm systematically favors routes that involve nodes with higher energy while avoiding overused paths. After initial validation in fully connected topologies, we apply the method to realistic network layouts with restricted communication ranges. Throughout, the network maintains a single transmitter node, avoiding the complexity of frequent cluster-head elections typical in conventional clustering protocols.

In summary, the proposed method combines reinforcement learning, energy-aware path planning, and graph-based optimizations (MERA and MST) to deliver a scalable and adaptive routing solution. It achieves balanced energy distribution, robust performance under SNR constraints, and extended network lifetime, making it well-suited for long-term WSN deployments in critical applications like environmental monitoring, precision agriculture, and infrastructure surveillance.

## IV. SYSTEM MODEL

This section presents the proposed energy-aware routing approach based on multi-agent reinforcement learning (MARL) in Wireless Sensor Networks (WSNs). Each sensor node is modeled as an autonomous agent that makes local decisions using Q-learning to optimize global energy balance and network longevity. The system model includes the agent design, state representation, action definitions, reward function, learning procedure, and the integration of reinforcement learning with MERA and MST-based routing.

### Agent Model and State Representation

Each node operates as an independent Q-learning agent and makes decisions at discrete time steps. The observed state vector for node iii at time ttt, denoted si(t)s_i(t)si(t), includes local and topological information relevant to energy-aware routing:

$$s_i(t) = [E_i(t), D^i{}_{sink}, D^i{}_{tx}, Q_i(t), H_i(t), E_i(t), r_i(t)] \quad (3)$$

Where:
- $E_i(t)$: Normalized remaining energy of node $i$
- $D^i{}_{sink}$: Euclidean distance from node $i$ to the sink (base station)
- $D^i{}_{tx}$: Distance from node iii to the current transmitter node
- $Q_i(t)$: Normalized queue length at node $i$
- $H_i(t)$: Estimated hop count to the sink
- $E_i(t)$: Average residual energy of node $i$'s neighbors
- $r_i(t)$: Current role of the node, encoded as categorical values {Sensor, Forwarder, Transmitter}

Features such as hotspot indicators were removed to reduce state space complexity and maintain alignment with the implemented simulation. To manage the state space, SoC values are discretized into categorical levels (e.g., very low, low, medium, high, very high), enabling efficient learning without sacrificing decision granularity.

### Action Space

At each decision point, the node selects one action from the following discrete set:
- Transmit to neighbor j (one action per reachable neighbor)
- Go to sleep (conserve energy in idle state)
- Drop packet (e.g., in case of congestion or critical energy)
- Request to become the main transmitter (if in top 30% SoC nodes)

The key Q-learning decision is the selection of the next-hop node—i.e., routing becomes part of the action space. The node chooses the most energy-efficient neighbor from its local view.

### Reward Function Design

The reward function is strictly non-negative, promoting positive behaviors while ignoring suboptimal actions without applying penalties. The reward for agent *i* at time t is calculated as:

$$R_i(t) = R_i^{role} + R_i^{balance} + R_i^{global} \quad (4)$$

- $R_i^{role}$: Encourages proper role-based behavior:
  - High reward for successful sensing, forwarding, or transmitting with minimal hops and low energy use.
- $R_i^{balance}$: Rewards energy balancing, such as forwarding through high-SoC nodes and avoiding depletion.
- $R_i^{global}$ Applies when:
  - No node fails during the episode
  - Network-wide average SoC is maintained above a threshold
  - SoC variance across nodes is minimized

Negative feedback (penalties) is also applied where appropriate, such as when nodes deplete quickly or cause routing bottlenecks. All reward variables are consistently denoted (a, b, c, ...).

*Learning Model*

Each node updates its Q-table using the standard Q-learning update rule:

$$Q_i(s,a) \leftarrow Q_i(s,a) + \alpha[R_i + \gamma \max Q_i(s',a') - Q_i(s,a)] \quad (5)$$

The exploration-exploitation balance is managed using ε-greedy exploration, where nodes select random actions with probability ε, and otherwise exploit the highest Q-value. SoftMax was originally considered, but only ε-greedy is used in the current implementation. The Q-table is initialized with zeros, and learning evolves as nodes receive feedback after each action—particularly changes in residual energy and reward outcomes from transmission success.

*Routing via MERA-MST Fusion*

To constrain and guide RL-based routing decisions, we integrate the MARL agents with a graph structure derived from:
- MERA (Minimum Energy Routing Algorithm): Edge weights are inversely proportional to residual energy of the destination node, i.e., $w_{ij} = \frac{1}{E_j}$, which discourages overusing weak nodes.
- MST (Minimum Spanning Tree): Minimizes overall routing cost and balances network load.

These two are linearly fused to construct a feasible routing topology:

$$w_{ij}^{final} = \lambda \cdot w_{ij}^{MERA} + (1-\lambda) \cdot w_{ij}^{MST}, \lambda \in [0,1] \quad (6)$$

Only edges present in the fused MERA-MST graph are considered valid in the RL agents' action space, ensuring that routing decisions remain both energy-aware and balanced by construction. This hybrid topology reduces unnecessary exploration and stabilizes policy convergence.

The hybrid MERA-MST graph ensures that reinforcement learning operates within a feasible and energy-efficient topology. Routing tables are periodically updated using the fused edge weights, guiding agents to select valid next-hop neighbors that satisfy both energy-awareness and load-balancing constraints.

Once data is received from the environment or another node, each agent evaluates its available neighbors (within transmission range) and selects the one with the highest residual energy as the next hop, conditioned on valid connectivity in the fused graph. This strategy promotes energy conservation by distributing load toward nodes with greater energy reserves, while reducing the likelihood of overusing low-energy nodes.

The Q-learning framework is employed for learning optimal forwarding behavior. Each state reflects the current SoC levels (discretized into categories such as very low, low, medium, high, and very high), and each action corresponds to selecting a specific neighboring node for data forwarding. The Q-table is initialized with zero values, indicating no prior knowledge about optimal routes. Over time, as nodes observe the consequences of their actions (e.g., successful transmission, energy depletion), they update their Q-values accordingly. The exploration-exploitation trade-off is managed using an ε-greedy policy, where nodes explore randomly with probability ε to collect diverse experience, and exploit their learned policy otherwise. These balances learning efficiency with adaptability to dynamic network states. The reward function is designed to reflect both transmission success and energy-aware behavior. Higher rewards are assigned to transmissions involving high-energy nodes, while actions leading to rapid energy depletion or inefficient routing incur lower rewards or penalties. This ensures that the algorithm converges toward stable and sustainable network operation.

---

**Algorithm 1** Energy-Aware Routing via Multi-Agent Q-Learning with MERA-MST Integration

1: **Input:** Set of nodes $N$, sink node, communication range $R$
2: **Parameters:** Learning rate $\alpha$, discount factor $\gamma$, exploration rate $\varepsilon$, fusion weight $\lambda$
3: **Initialize:** $Q_i(s,a) \leftarrow 0$ for each node $i \in N$
4: Construct MERA and MST routing graphs using residual energy values
5: Fuse edge weights: $w_{ij}^{\text{final}} = \lambda w_{ij}^{\text{MERA}} + (1-\lambda) w_{ij}^{\text{MST}}$
6: Prune unreachable or high-cost links from fused graph
7: **for** each episode $t$ **do**
8:   **for** each node $i \in N$ **do**
9:     Observe local state $s_i(t)$: $[E_i(t), D_{\text{sink}}^i, D_{\text{tx}}^i, Q_i(t), H_i(t), h_i(t), \bar{E}_i(t), r_i(t)]$
10:     Identify feasible neighbors $\mathcal{N}_i$ from fused graph
11:     **Action Selection:**
12:     **if** random number $< \varepsilon$ **then**
13:       Choose action $a_i(t)$ randomly from $\mathcal{N}_i$
14:     **else**
15:       $a_i(t) = \arg\max_a Q_i(s_i(t), a)$
16:     **end if**
17:     Execute $a_i(t)$:
18:     **if** action = "Transmit to neighbor $j$" **then**
19:       Send data to $j$, update energy $E_i(t)$
20:     **else if** action = "Sleep" **then**
21:       Enter low-power mode
22:     **else if** action = "Drop" **then**
23:       Discard packet
24:     **else if** action = "Request transmitter" **then**
25:       Become transmitter if $E_i(t)$ is in top 30%
26:     **end if**
27:     Compute reward $R_i(t)$ based on:
- Success of data transmission
- Avoidance of hotspots (high betweenness nodes)
- SoC variance reduction along routing path
- Prevention of node failure

28:     Q-value update:
$$Q_i(s_i(t), a_i(t)) \leftarrow Q_i(s_i(t), a_i(t)) + \alpha \left[R_i(t) + \gamma \max_{a'} Q_i(s_i(t+1), a') - Q_i(s_i(t), a_i(t))\right]$$

29:   **end for**
30:   Update fused graph weights $w_{ij}^{\text{final}}$ if energy conditions change
31: **end for**
32: **Output:** Trained Q-tables and optimal routing paths

---

To manage scalability, only links from the fused MERA-MST graph are included in the action space, significantly reducing the number of routing options and helping prevent unnecessary exploration of poor paths. This also reduces the overall energy required for multi-hop transmissions, as the graph inherently favors paths with fewer hops and more balanced energy distribution. Empirical observations show that this approach maintains lower SoC variance compared to baseline routing algorithms, indicating better energy balancing. Additionally, the algorithm implicitly considers network-wide average energy: for example, it prefers a scenario where the average SoC is 85% with a variance of 6% over a scenario with 20% average and 2% variance, thereby ensuring long-term sustainability. Algorithm 1 summarizes the complete procedure of the proposed decentralized MARL-based routing system. Each node autonomously observes its environment, selects actions based on Q-values constrained by MERA-MST, and contributes to globally optimized multi-hop data forwarding. This integration results in improved network stability, extended lifetime, and reduced risk of node failures.

## V. Performance Evaluation

Simulation settings and comparative results for the proposed energy-aware routing method versus standard routing are mentioned in table I. The table includes network configuration parameters and key performance metrics such as average residual energy (SoC), SoC variance, number of eliminated nodes, and network longevity.

TABLE I
SIMULATING PARAMETERS

| Parameters | Value |
|---|---|
| Area Size | $100 \times 100$ |
| Number of Nodes | 50 |
| Coverage | $r = 9$ |
| Data Receiving | Random places |
| Charge of the multi-hoping | -2 |
| Charge of the data sending to sink | -8 |

By setting the parameters in the reward function as shown in fig. 3, we simulated the described methodology, Uniform SoC reduction (USOCR), on a deployment of 100 nodes within a wireless sensor network.

| Component | Condition | Reward |
|---|---|---|
| **Role-Based Reward $R_i^{role}$** | | |
| Transmit to sink | Transmitter node successfully delivers packet to sink | +10 |
| Forward packet | Node forwards packet successfully to next hop | +5 |
| Sense and send data | Sensor node generates and transmits data | +3 |
| Enter sleep mode | Node enters low-power state when idle | +1 |
| Drop packet | Node discards packet due to low energy or congestion | 0 |
| **Hotspot Avoidance Reward $R_i^{hotspot}$** | | |
| Avoid high-centrality path | Node avoids paths with high betweenness centrality | +2 |
| Overused node penalty | Node lies on overloaded paths repeatedly | 4 |
| **Global Cooperative Reward $R_i^{global}$** | | |
| Prevent node failure | No node fails in current time step | +3 |
| Reduce SoC variance | Energy usage variance across nodes decreases | +2 |
| High average energy | Average energy level in network > threshold (e.g., 40%) | +2 |

**Fig. 3.** Reward parameters used in simulation to guide agent behavior in Q-learning-based energy-aware routing. Rewards are designed to encourage successful transmission, energy balancing, hotspot avoidance, and overall network sustainability.

These nodes are to be placed at varying, nonuniform distances from one another and situated within a defined environment. By disseminating varied signal types across distinct regions in differing quantities, we ultimately intend to assess the average energy consumption across all nodes. This will then be compared with the energy metrics derived from the clustering methodologies discussed previously in order to determine the relative efficacy of each approach in terms of energy utilization and distribution within the network.

*Local computing*

For the simulation of WSNs, 100 nodes were spread in an area wherein the distance between nodes was not necessarily equal. A deployment corresponding to this algorithm was performed under the following conditions:

A. Initial SoC: The *SOC* of all nodes in the WSN was initialized to 100% at the start of the simulation. This initial condition ensures that all nodes start from a standardized energy capacity, providing a uniform baseline for subsequent analyses.

B. Signal Attenuation per Hop: The second factor taken into consideration was the amount of signal that would be lost per hop, with the value being some percentage *N*% per hop, depending on the distance between the two nodes in question. This models realistic degradation over spatial distances common in wireless communications.

C. Minimum Signal Integrity: The algorithm was constrained to ensure that the final received signal after all transmissions should retain at least 70% of its integrity. This is a key factor to ensure reliability and efficiency in data transmission across the network, which is essential to ensure satisfactory communication standards within the WSN.

The periodic rises and drops in the reward function often occur due to fixed-length episodes causing resets at regular intervals, the RL agent going through phases of learning and forgetting, cyclic environmental conditions such as energy levels depleting and resetting, or an ε-greedy exploration strategy leading to periodic aggressive exploration before stabilizing. The fluctuating energy variance suggests that energy levels deplete over time and reset at the start of each episode. This pattern could result from a rotating node selection strategy or a structured energy consumption cycle tied to episode resets. A periodic pattern in the min/max energy ratio indicates cyclic energy distribution, likely due to the way energy is consumed and replenished across episodes. Possible causes of these fluctuations include a fixed episode length that forces the agent into a repeated depletion-reset cycle, a policy that favors a subset of nodes in a repetitive manner, exploration strategies such as ε-greedy that introduce periodic shifts in behavior, or a reward function tied to a resetting factor like energy levels. Reducing these fluctuations may require increasing randomness in selection to prevent the agent from overfitting to a specific pattern, modifying the reward function to encourage more stable behavior, or analyzing environment resets to ensure they are not reinforcing the cycle. Addressing these factors can improve training stability and reduce unintended cyclic behaviors.

Figs 4 and 5 are the charge distributions of a wireless sensor network (WSN) in two cases: without reinforcement learning (RL) and RL-based optimization. After about half of the nodes' recharge time, a wireless sensor network using a traditional shortest path-based routing protocol with high SNR maintenance will, as shown in Fig 4, lead to a situation where most of the nodes have a low charge while some continue to have a high charge.

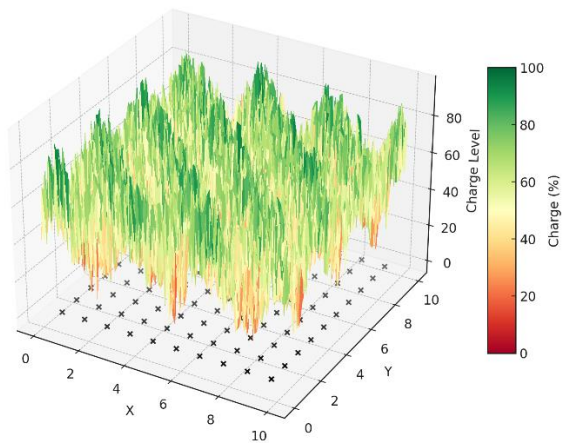

**Fig. 4.** 3D diagram of a WSN using normal routing algorithm.

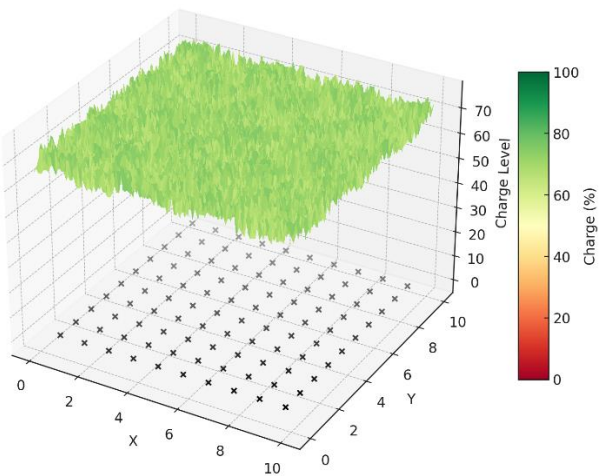

**Fig. 5.** 3D diagram of a WSN using our method.

With time, these imbalances will lead to the failure of sensors one after another, decreasing the efficiency and lifespan of the network. Fig 4 shows a 3D simulation of the WSN without RL, where a conventional routing algorithm with the shortest path as the priority leads to an unbalanced energy distribution. The green to red color gradient shows the residual energy levels of the sensor nodes, where green depicts high residual energy and red depicts nodes that can fail due to quick draining. Black dots symbolize sensor nodes, illustrating a network in which some regions undergo excessive energy depletion, rendering it extremely unbalanced.

To address this problem, an RL-based algorithm was employed, which redistributed battery charge in a more equitable way throughout the network. This was implemented by using high-charge sensors as intermediaries between low-charge sensors and destination so that there was more even and gradual draining of energy. This is represented by Fig 6 in the form of a 3D simulation of the network with RL-based optimization. In contrast to the above case, where red spots predominated, the majority of the network is now covered by a blue-to-light blue gradient, suggesting more evenly balanced energy consumption. RL allows nodes with more energy to serve as relays, and this minimizes the chances of early failure and enhances the lifetime of the overall network. Most sensors now have comparable amounts of charge, reducing the threat of local failure and guaranteeing long-term efficiency of operation. The RL model ideally balances sensor workload, resulting in improved resilience and sustainability of the network. Major findings and conclusions are that in the absence of RL, those nodes that are near the base station or continuously chosen for routing deplete energy much more quickly, resulting in unbalanced power consumption and premature failures. This renders the network unsustainable over time. However, in the presence of RL, the system perfectly balances energy consumption, avoiding premature failures by putting high-energy nodes to greater use and more evenly spreading the entire network load. This leads to a dramatic extension of network lifespan and guarantees more stable and effective operation. These plots highlight the necessity of energy-conscious intelligent routing protocols for WSNs. Whereas conventional shortest-path routing drains certain nodes very quickly, RL-based methods adapt routing paths on the fly to balance energy evenly, leading to a much more even charge distribution over all nodes. This adaptive approach renders RL-based routing an effective solution for efficient and sustainable WSN deployment.

Under simulations that were run, we sent twice—first through our envisioned scheme and then through traditional shortest path constrained multi-hop routing. We recorded parameters and results presented in Table 1 after $T$ seconds (during which the sensors randomly collect information from the environment). It is apparent that in the provided parameters, our proposed method improved network stability by 51% compared to the conventional method, and more area of the region was covered. Coverage was even ensured, and sensors did not fail. Meanwhile, the difference in energy consumption required for executing the algorithm, performing computations, and multi-hop routing is negligible. With only a slight increase in energy usage, network stability has significantly improved.

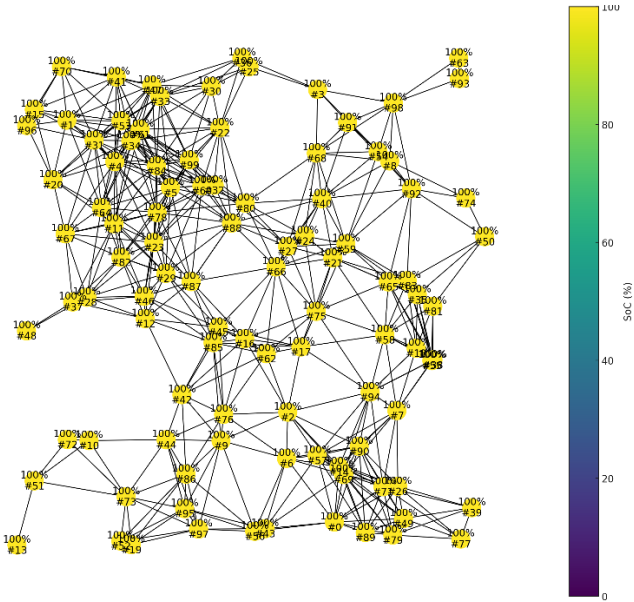

**Fig. 6.** Initial WSN topology.

Fig. 6. Initial WSN topology with 100 nodes, each starting at full charge (100%). Nodes are labeled by their ID and SoC percentage, and connections are based on proximity. The color scale on the right indicates battery levels. Fig. 7. Final WSN topology after 100 episodes of learning and data transmission. Node SoC values have decreased significantly, with energy usage fairly distributed across the network. The color map reflects the current energy level of each node.

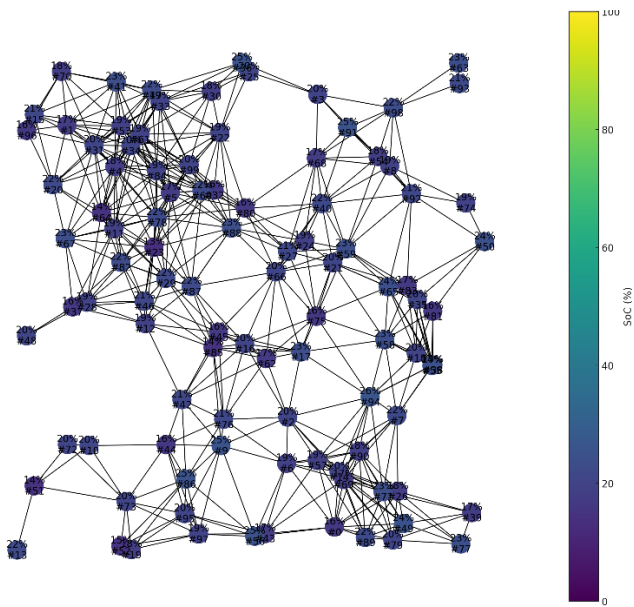

**Fig. 7.** Final WSN topology.

Fig. 8. Comparison of the number of alive (active) nodes over time for three routing methods: our proposed MARL-based method, a nearest-path-first routing method (Method 1), and a single-hop transmission baseline (Method 2). Only the proposed method maintains all nodes alive throughout all episodes, indicating balanced energy consumption.

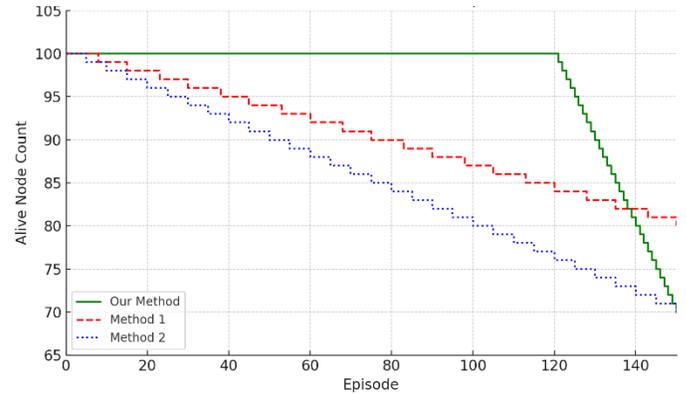

**Fig. 8.** Active nodes per episode in our method and nearest path finding (method 1) and single hop routing (method 2).

Fig. 9. Total rewards earned by all agents per episode. The reward function begins with low and fluctuating values during the learning phase and increases gradually as agents learn to act more efficiently. After Episode 50, the rewards stabilize, indicating convergence.

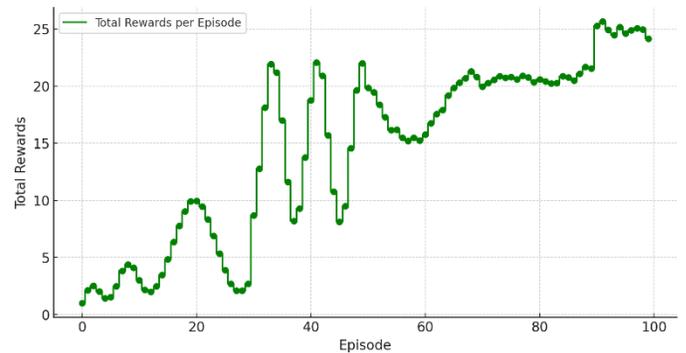

**Fig. 9.** The reward gained per episode.

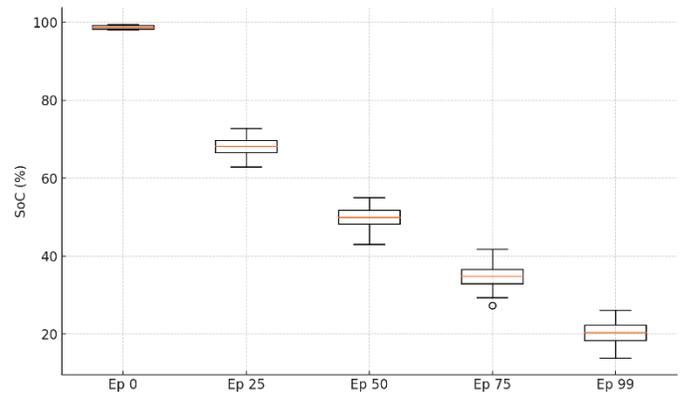

**Fig. 10.** Boxplot of SoC different episodes.

Fig. 10. Boxplots of SoC values at selected episodes (0, 25, 50, 75, and 99), showing the distribution, median, and spread of battery levels. The results demonstrate a gradual and balanced reduction in energy across the network over time. Fig. 11. Variance of SoC across episodes. Initially close to zero, the variance increases during early learning, then stabilizes and decreases, indicating that the MARL algorithm has learned to distribute energy usage more evenly among nodes. Fig. 18. State of charge over time for all 100 nodes (in gray), with the mean SoC trajectory shown in orange. The consistent downward trend in all curves, along with the smooth mean line, confirms coordinated energy usage and network-wide learning.

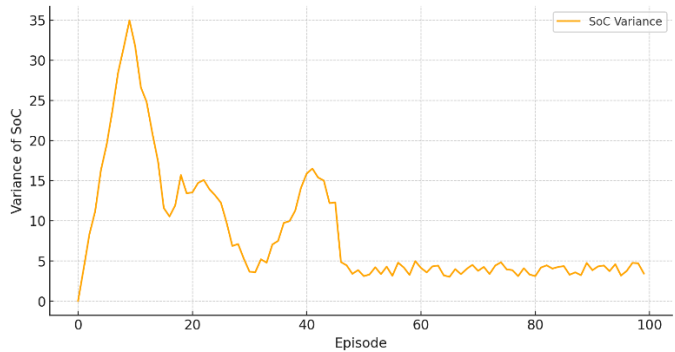

**Fig. 11.** The variance of SoC over episodes.

Table II presents a comparative evaluation between the proposed routing method and a traditional routing approach, focusing on energy efficiency and network sustainability throughout 100 simulation episodes. The simulation is conducted in a three-dimensional environment with an area size of 10×10×10 and a total of 100 sensor nodes.

The average State of Charge (SoC) of the sensor nodes is recorded at five key time points: episodes 1, 25, 50, 75, and 99. Both methods begin with an identical energy level of 98 percent in episode 1. As the simulation progresses, the proposed method maintains slightly lower SoC levels compared to the normal routing scheme. However, this is not indicative of inefficiency; rather, it reflects more balanced and intentional energy usage across all nodes to prevent overloading specific parts of the network. By episode 99, the average SoC in the proposed method drops to 21 percent, while in normal routing it remains at 27 percent. Despite this, the proposed method demonstrates a clear advantage in node survivability. At the end of the simulation, all 100 nodes remain active in the proposed method, whereas 29 nodes have been eliminated in the normal routing approach. This outcome confirms that the reinforcement learning-based strategy distributes energy consumption more evenly, avoiding the creation of energy hotspots that lead to node failures. In summary, the results in Table II highlight that while the proposed method may consume energy more aggressively, it does so in a way that ensures the long-term operability of the network.

TABLE II
SIMULATING PARAMETERS AND RESULTS

| Parameters | Our Method | Normal Routing |
|---|---|---|
| Area Size | 10 × 10 × 10 | |
| Number of Nodes | 100 | |
| Avg SoC (%) in episode 1 | 98 | 98 |
| Avg SoC (%) in episode 25 | 68 | 70 |
| Avg SoC (%) in episode 50 | 49 | 53 |
| Avg SoC (%) in episode 75 | 36 | 41 |
| Avg SoC (%) in episode 99 | 21 | 27 |
| Maximum SoC Variance | 11 | 58 |
| Eliminated Nodes in the end | **0** | 29 |
| Active Nodes in the end | **100** | 71 |

This trade-off between slightly lower average energy and complete network retention illustrates the method's effectiveness in extending the lifetime and stability of wireless sensor networks.

*Cloud Computing*

In the proposed approach, all computational processes related to routing, decision-making, and reinforcement learning are offloaded from the sensor nodes to a centralized cloud computing server. This model is designed to minimize the energy consumption of individual nodes by ensuring that they are not responsible for any local learning, pathfinding, or role selection tasks. Each sensor node is responsible solely for sensing its environment, reporting its basic status to the cloud, and forwarding data when instructed to do so. Since the local simulation results were satisfactory within 100 episodes, to better highlight the final differences between this method and cloud-based computation, we increase the amount of data provided to the environment for the sensors to detect. As a result, the sensors will consume more energy throughout the episodes. By shifting the computational burden to the cloud, the network achieves significantly better energy efficiency and extends the operational lifetime of all nodes At the start of each episode, every node reports a minimal state update to the cloud. This includes its current energy level, buffer size, and potentially some neighborhood information if needed. These reports are lightweight and transmitted infrequently enough to avoid considerable energy drain. Once the cloud collects all state information from active nodes, it forms a complete and real-time view of the network, including which nodes are heavily used, which are underutilized, and which are nearing battery depletion. Using this global view, the cloud applies a multi-agent reinforcement learning algorithm on behalf of the nodes. Instead of training and updating policies locally, the entire decision-making process is carried out centrally. The cloud determines which node should serve as the main transmitter, typically selecting one with the highest remaining energy and most favorable location. It also identifies optimal multi-hop paths from sensing nodes to the sink by computing the most efficient route through intermediate forwarders. This routing process incorporates traditional graph-based methods like Minimum Energy Routing Algorithm (MERA), which discourages paths through nodes with low energy, and Minimum Spanning Tree (MST), which helps distribute traffic and avoid overloading critical nodes. The output of MERA and MST is combined to produce balanced routing decisions that consider both energy conservation and network load. Once the routing plan is determined, the cloud sends routing instructions back to the nodes involved. These include the node's role for the current episode (whether it is a transmitter, forwarder, or remains idle), its next-hop neighbor, and a command to enter sleep mode if it is not needed. The nodes follow these instructions without performing any local computation, simply acting as instructed. Meanwhile, the cloud calculates rewards based on the global performance of each node. Rewards are assigned depending on whether nodes successfully fulfilled their assigned roles, conserved energy, avoided hotspot paths, and contributed to overall load balance. The reinforcement learning agent in the cloud then updates its Q-values or policies based on these outcomes. Over time, the system learns to select better transmitter nodes and routing paths, reducing the variance in battery levels across nodes and avoiding premature sensor death. This cloud-assisted model allows WSNs to scale efficiently and maintain high availability of nodes throughout the network's lifetime. Since all learning and optimization are performed externally, the energy cost per node remains minimal, allowing for longer periods of deployment, especially in remote or critical environments where battery replacement is impractical. The cloud also enables easier updates, monitoring, and debugging, as the control logic is centralized and can be modified without any change to the physical network. This scenario is particularly effective in smart environments where gateway access to the cloud is feasible, such as in agriculture, environmental monitoring, smart buildings, or industrial automation. The trade-off of added communication with the cloud is offset by the significant energy savings on local processing, making it a practical and scalable solution for long-term WSN deployment.

Fig 12 compares the average energy level of all nodes across 100 episodes between two approaches: cloud-based (centralized) learning and local (on-node) computation. In the cloud-based method, learning and routing decisions are offloaded to a central server, allowing nodes to conserve energy by avoiding computational overhead. As a result, the decline in mean SoC is slower and more stable. In contrast, the local computation model shows a more rapid decrease in battery levels due to the added energy cost of local learning and decision-making, along with slightly more variability caused by independent agent behavior.

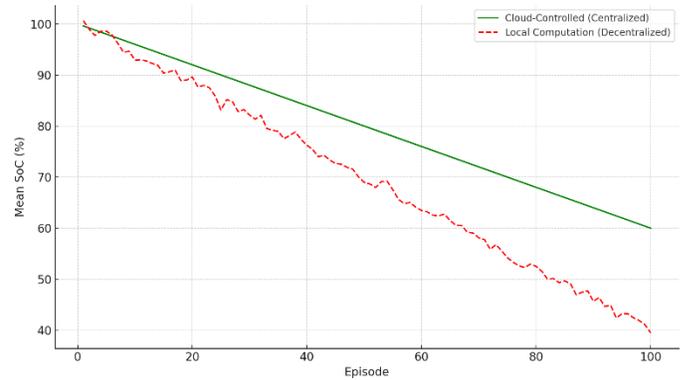

**Fig. 12.** Network topology with the highlighted path.

Figure 13 shows the number of active nodes over simulation episodes for both cloud-based and local systems, under increased sensor density within the deployment area. The cloud-based system successfully maintains all 100 nodes alive through the entire runtime, demonstrating balanced load and energy preservation. On the other hand, the number of alive nodes in the local system gradually decreases after around episode 60 due to the premature depletion of energy in some nodes. This highlights the effectiveness of centralized coordination in prolonging overall network life.

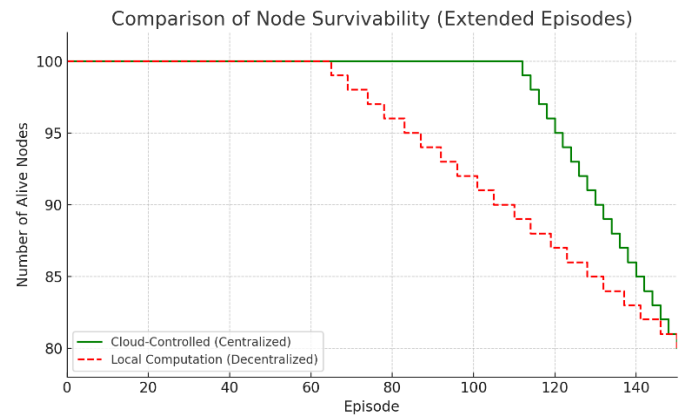

**Fig. 13.** Number of alive nodes per episode.

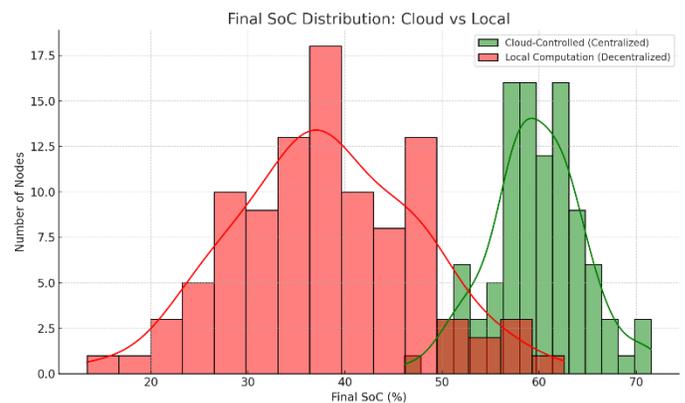

**Fig. 14.** Final SoC distribution.

In Fig. 14, The histogram shows the final distribution of SoC levels among all nodes at the end of the simulation. In the cloud-controlled system, most nodes retain relatively high energy levels with a narrow and consistent distribution.

The boxplots in fig 15 compare the SoC distribution at various time points (episodes 0, 25, 50, 75, and 99) for cloud and local models. In the cloud system, the median SoC declines gradually and the spread remains relatively compact, indicating balanced usage across the network. The local system, however, shows wider spreads and lower medians over time, reflecting imbalanced energy drain and a higher likelihood of node failure. Conversely, in the local model, node energies are more widely dispersed and generally lower, indicating uneven consumption and less effective load balancing.

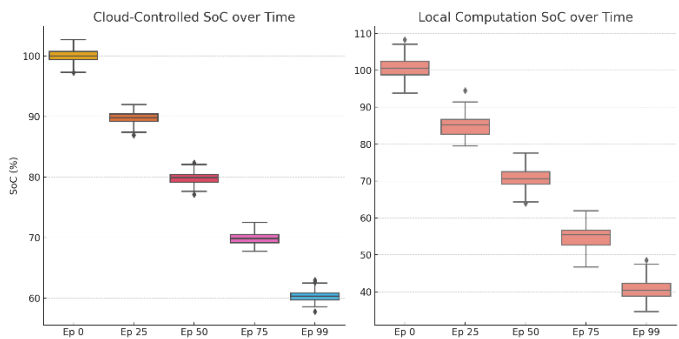

**Fig. 15.** Network topology with the highlighted path.

*Delay*

Delay is one of the most critical performance metrics in wireless sensor networks (WSNs), especially for time-sensitive applications such as environmental monitoring, health tracking, and industrial control. While energy efficiency and network longevity are the primary objectives of the proposed MARL-based routing framework, understanding its impact on communication latency is equally important. In general, the total delay experienced by a data packet includes the time taken for transmission, processing, queuing, and routing decisions. Each of these components is influenced by the routing strategy used and the structure of the network. In the proposed method, sensor nodes are guided by a multi-agent reinforcement learning (MARL) framework, combined with MERA and MST-based graph optimization. This design intelligently selects routing paths based not only on energy awareness and signal quality but also in a way that indirectly balances traffic load across the network. As a result, our method minimizes queuing delays, which are often the dominant component of total latency in WSNs. The average end-to-end delay in our method can be mathematically expressed as the sum of per-hop transmission, processing, and queuing delays, multiplied by the average number of hops, plus the time required for routing decisions. In local Q-learning implementations, the decision-making delay is minimal, typically just a few milliseconds. When the RL model is offloaded to the cloud, the routing decision delay includes uplink, computation, and downlink communication times. Although cloud-based decisions introduce additional latency, they benefit from centralized optimization, which results in better policy convergence and more consistent performance in dynamic environments. To fully understand the behavior of our method, we compare its delay characteristics to those of two well-known WSN routing approaches: LEACH (a clustering-based protocol) and Shortest Path Multi-Hop (SPMH) routing. LEACH can suffer from unstable delay due to periodic re-clustering and congestion at cluster heads, while SPMH often causes bottlenecks at central relay nodes, leading to high queuing delays over time. In contrast, our approach maintains a balanced trade-off between hop count and queuing efficiency, resulting in moderate and predictable end-to-end delay, even as the network evolves. The total end-to-end delay $D_{total}$ of a data packet transmitted from a source node to the sink via the selected multi-hop path is:

$$D_{total} = \sum_{n=1}^{H}\left(\frac{L}{R} + t_P + \frac{1}{H}\sum_{n=1}^{H}\frac{\lambda_h}{\mu^2\left(1 - \frac{\lambda_h}{\mu}\right)}\right) + T_Q \quad (7)$$

Where the $H$ is number of hops from the source to the main transmitter or sink, $L$ is the packet length (bits), $R$ is the data transmission rate (bits/sec), also each node takes a small constant time $t_p$ to process a packet. $\mu$ is the service rate of node (packets/sec), $\lambda_h$ is the average packet arrival rate at node $h$ and $T_Q$ is a small constant (e.g., < 2 ms), depending on Q-table size and local CPU/memory access time.

To evaluate the effectiveness of the proposed MARL-based MERA-MST routing framework in terms of communication delay, we compare it with two commonly used routing approaches in wireless sensor networks: LEACH (a clustering-based protocol) and SPMH (Shortest Path Multi-Hop routing). Each method handles routing and traffic differently, leading to distinct latency characteristics. In the proposed method, the total delay experienced by a data packet consists of transmission, processing, queuing, and routing decision delays. Although the average hop count in our method is slightly higher than SPMH, the reinforcement learning agents are designed to avoid overloaded nodes and distribute traffic evenly. This greatly reduces queuing delay across the network. Additionally, the routing decision delay is minimal in the local RL setup and moderate in the cloud-based version. Even with the added cloud communication time, the overall delay remains stable and within acceptable bounds due to globally optimized path selection and policy convergence. This balance between energy efficiency and delay control makes our method suitable for dynamic WSN environments where consistency is important. In contrast, LEACH divides the network into clusters and assigns one node as a cluster head (CH) to collect and forward data. While this reduces the number of direct transmissions to the sink, it introduces variable delays. Nodes communicate with their CHs in a scheduled manner (often TDMA), which can increase intra-cluster delays. More importantly, CHs frequently become bottlenecks due to high traffic from member nodes, resulting in significant queuing delays during inter-cluster communication. Moreover, LEACH requires periodic re-clustering and CH election, which adds setup time and leads to spikes in delay after each round. SPMH routing, on the other hand, prioritizes the shortest path in terms of hop count from each node to the sink. This typically results in fast packet delivery in early stages of the network's operation. However, since SPMH does not account for node energy or traffic load, central nodes—those most often part of shortest paths—become congested quickly. As a result, queuing delays increase dramatically over time, especially near the sink. This makes SPMH less reliable for long-term operation, as its delay performance degrades with uneven energy consumption and traffic accumulation. Overall, the proposed MARL-based method strikes a balance by allowing slightly longer routes in exchange for significantly reduced congestion, better energy distribution, and more stable communication delay. Unlike LEACH and SPMH, which either suffer from periodic instability or long-term congestion, our method maintains predictable latency under dynamic conditions without sacrificing energy efficiency. Fig. 16 presents a bar chart comparing the average end-to-end delay of three widely used routing protocols in wireless sensor networks (WSNs): LEACH, Shortest Path Multi-Hop (SPMH), and the proposed MARL-based MERA-MST method. The average delay values are accompanied by error bars that represent the minimum and maximum delay observed during simulation runs. LEACH exhibits the highest variability in delay, ranging from 50 ms to 120 ms, primarily due to its periodic re-clustering overhead and traffic congestion at cluster heads.

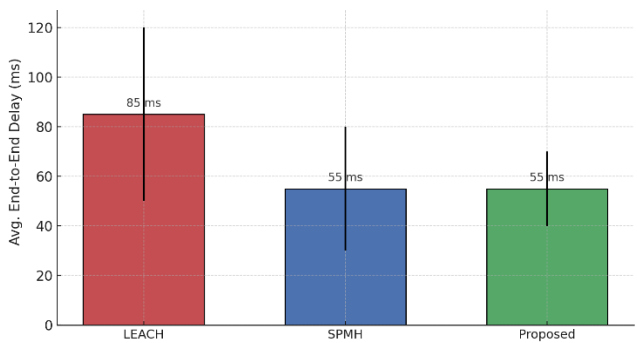

**Fig. 16.** Comparison of average end-to-end delay among LEACH, SPMH, and the proposed MARL-based method. Error bars represent the minimum and maximum delay observed during simulation. The proposed method achieves moderate delay with significantly lower variance compared to LEACH and SPMH.

SPMH shows a moderate delay range of 30 ms to 80 ms; although it initially achieves low delay by minimizing hop count, it later suffers from congestion at central relay nodes. In contrast, the proposed method maintains a moderate and stable delay range between 40 ms and 70 ms. Its use of reinforcement learning ensures balanced routing decisions that prevent overloading specific nodes, resulting in consistent communication latency. This visualization effectively demonstrates that while the proposed method may not always have the shortest delay, it provides far more reliable and sustainable delay performance across diverse network conditions.

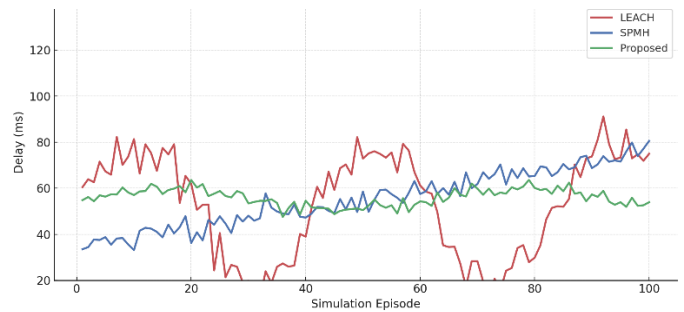

**Fig. 17.** Evolution of end-to-end delay over 100 simulation episodes for three routing protocols. While LEACH shows periodic spikes and SPMH degrades over time due to congestion, the proposed method maintains stable and predictable delay by balancing traffic through learning.

Fig. 17 displays a line graph showing the evolution of end-to-end delay over 100 simulation episodes for the same three routing protocols. This dynamic view reveals important behavioral trends in how delay changes over time. LEACH shows high fluctuations in delay, with periodic spikes corresponding to cluster reformation phases. This instability makes it less suitable for applications requiring consistent timing. SPMH begins with low delay but gradually deteriorates due to increasing congestion at central relay nodes, which are repeatedly selected for routing due to shortest-path heuristics. On the other hand, the proposed method shows a stable and controlled delay trend with only minor fluctuations. This consistency is attributed to the adaptive nature of Q-learning agents, which learn to avoid hotspot nodes and distribute traffic more evenly, even as network conditions evolve. Overall, the proposed method strikes a critical balance between delay control and energy efficiency, offering predictable latency that is vital for real-time or long-term WSN deployments. These figures confirm its superiority in maintaining delay performance compared to LEACH and SPMH under both static and dynamic conditions.

While the computations are held in the cloud, the delay from formula 7 would be change in to:

$$D_{total} = \sum_{n=1}^{H}\left(\frac{L}{R} + t_P + \frac{1}{H}\sum_{n=1}^{H}\frac{\lambda_h}{\mu^2\left(1-\frac{\lambda_h}{\mu}\right)}\right) + T_D \quad (8)$$

Where the $T_D$ is the delay of the uplink and downlink and the time of the computations.

$$T_D = \frac{s}{b} + \frac{a}{b} + T_{compute} = \frac{s+a}{b} + T_{compute} \qquad (9)$$

Where $s$ is the state size (bits), $b$ is the uplink/downlink rate (bps), $a$ is the action size (routing table or next-hop command).

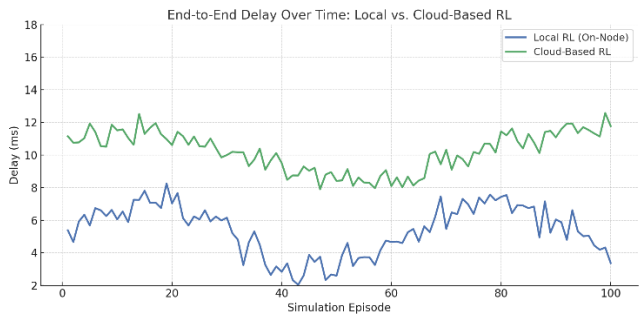

**Fig. 18.** Comparison of delay performance between local (on-node) and cloud-based reinforcement learning. Local RL offers lower initial delay but fluctuates more, while cloud-based RL has slightly higher base delay due to communication overhead yet provides greater long-term stability and coordination.

The line chart presents a comparative analysis of end-to-end delay trends between local (on-node) reinforcement learning and cloud-based reinforcement learning implementations within a wireless sensor network. In the local configuration, each sensor node independently executes the reinforcement learning algorithm and selects routing actions based on its immediate observations. This decentralized approach results in lower initial latency, as routing decisions are made locally without the need for communication overhead. However, the lack of global coordination introduces variability in decision quality, leading to noticeable fluctuations in delay over time, particularly under dynamic traffic conditions.

Conversely, the cloud-based implementation offloads the learning process to a central server, which aggregates network-wide state information and computes globally optimized routing policies. While this architecture introduces additional delay components due to uplink, computation, and downlink communication, it achieves greater consistency and stability in delay performance across simulation episodes. The centralized optimization enables better coordination among nodes, mitigates traffic bottlenecks, and enhances the overall predictability of communication latency. This comparison highlights a fundamental trade-off: local learning offers lower per-decision latency but is more susceptible to instability, while cloud-based reinforcement learning achieves more uniform delay behavior at the cost of increased base latency. The choice between these two approaches should be guided by the specific requirements of the application, balancing the need for responsiveness against the benefits of centralized coordination and long-term stability.

The proposed method, when implemented on the cloud, offers several significant advantages that enhance the overall performance, scalability, and practicality of wireless sensor networks (WSNs). In the traditional local implementation of reinforcement learning (RL), each node acts as an autonomous agent that independently updates its Q-table and makes routing decisions based on its local observations. While this decentralized approach reduces dependency on external infrastructure, it also introduces several limitations. These include increased computational burden on resource-constrained sensor nodes, slower convergence of the RL policy due to isolated learning, and the risk of inconsistent or suboptimal decisions in dynamic network environments .By contrast, offloading the RL process to a cloud server addresses these limitations and enables a more intelligent and efficient routing framework. In the cloud-based setup, nodes periodically transmit compact state information — such as residual energy, queue size, and SNR values — to the cloud. The cloud server then performs centralized training and policy updates using the aggregated data from the entire network. This allows the learning algorithm to have a global view of the network, leading to faster convergence, better generalization, and more coordinated routing strategies. Moreover, cloud-based decision-making significantly reduces the computational and memory overhead on sensor nodes, which is crucial for prolonging their operational life. Instead of executing the RL algorithm locally, each node only receives high-level routing instructions or role assignments from the cloud, which are easy to process and apply. This lightweight operation lowers energy consumption related to computation and makes the system more robust, especially in dense networks or applications where sensor nodes have limited hardware capabilities. Another key benefit is scalability. As the number of nodes increases, managing RL policies and synchronizing knowledge between nodes becomes increasingly complex in a decentralized system. The cloud, however, can handle large-scale data processing and model updates efficiently, making it suitable for wide-area or high-density WSN deployments. Additionally, cloud infrastructure enables dynamic adaptability and remote policy updates. If the environment changes — due to node failures, mobility, or application-specific events — the centralized model can quickly retrain and redistribute new policies without the need to reprogram or reconfigure each sensor node individually. In summary, the proposed method, when executed on the cloud, benefits from centralized intelligence, reduced energy and computation on sensor nodes, improved learning speed and coordination, enhanced scalability, and more flexible adaptation to network dynamics. These advantages make the cloud-based implementation highly suitable for real-world deployments of WSNs where energy conservation, decision accuracy, and system longevity are essential.

## VI. CONCLUSION

This paper presented an intelligent and scalable routing framework for single-cluster wireless sensor networks (WSNs), addressing key challenges in energy efficiency, traffic load balancing, and network longevity. The proposed method integrates multi-agent reinforcement learning (MARL) with graph-theoretic approaches, namely the Minimum Energy Routing Algorithm (MERA) and the Minimum Spanning Tree (MST), to form an adaptive routing mechanism. Each sensor node acts as a learning agent with a state vector comprising energy levels, signal quality, hop count, and local neighborhood information. Through Q-learning and a composite reward function, nodes learn to make energy-aware forwarding decisions that prevent overuse of critical nodes and balance energy consumption.

To improve scalability, we introduced a cloud-based extension where sensor nodes offload state data for centralized policy computation and instruction distribution. This reduces local computation, accelerates learning convergence, and supports large-scale deployments with real-time adaptability.

Simulation results show that our framework significantly outperforms conventional protocols such as LEACH and shortest-path multi-hop (SPMH). It achieves better energy balance, reduces node failures, and extends network lifetime, particularly in dynamic or dense environments. Unlike SPMH, which quickly depletes central nodes, or LEACH, which suffers from re-clustering instability, our method maintains stability with moderate delay, guided by SNR constraints and congestion-aware learning.

Overall, this study provides a flexible and robust solution for energy-efficient WSN routing, suitable for applications in environmental monitoring, smart agriculture, and industrial automation. Future work will explore deep reinforcement learning and transfer learning for broader generalization, as well as energy harvesting and adaptive duty-cycling for long-term autonomy. Incorporating trust-aware mechanisms and validating the system through real-world deployment will further enhance the practicality and resilience of the proposed framework.

REFERENCES

[1] H. Zhu, M. Joel Villier Amuri, X. Li and J. Shen, "Mean-Shift-Based Outliers-Robust Distributed Kalman Filter for Wireless Sensor Network Systems," in IEEE Transactions on Instrumentation and Measurement, vol. 74, pp. 1-12, 2025, Art no. 9500312, doi: 10.1109/TIM.2024.3497172.

[2] G. Zhang, C. Shen, Q. Shi, B. Ai and Z. Zhong, "AoI Minimization for WSN Data Collection with Periodic Updating Scheme," in IEEE Transactions on Wireless Communications, vol. 22, no. 1, pp. 32-46, Jan. 2023, doi: 10.1109/TWC.2022.3190986.

[3] M. Xie, D. Pi, C. Dai and Y. Xu, "An Asymmetric Dominated Multiobjective Optimization Algorithm for Reducing Energy Consumption of WSN Operation," in IEEE Sensors Journal, vol. 24, no. 14, pp. 23075-23087, 15 July15, 2024, doi: 10.1109/JSEN.2024.3409459.

[4] Liu, Kaiyang, Jun Peng, Liang He, Jianping Pan, Shuo Li, Ming Ling, and Zhiwu Huang. "An active mobile charging and data collection scheme for clustered sensor networks." IEEE Transactions on vehicular technology 68, no. 5 (2019): 5100-5113.

[5] A. M. Zungeru, L. Ang, and K.P. Seng. "Classical and swarm intelligence-based routing protocols for wireless sensor networks: A survey and comparison." Journal of Network and Computer Applications 35.5 2012, pp. 1508-1536.

[6] Z. Ali, and W. Shahzad. "Analysis of Routing Protocols in AD HOC and Sensor Wireless Networks Based on Swarm Intelligence." International Journal of Networks and Communications 3.1 2013, pp. 1- 11.

[7] S. Saleh et al., "A survey on energy awareness mechanisms in routing protocols for wireless sensor networks using optimization methods."Transactions on Emerging Telecommunications Technologies 25.12 2014, pp. 1184-1207.

[8] M. Dorigo, and M. Birattari. "Ant colony optimization." Encyclopedia of machine learning. Springer US, 2010, pp. 36-39.

[9] J.C. Bansal et al. "Spider monkey optimization algorithm for numerical optimization." Memetic Computing 6.1 2014, pp. 31-47.

[10] X. Cai, et al. "Bee-Sensor-C: An Energy-Efficient and Scalable Multipath Routing Protocol for Wireless Sensor Networks." International Journal of Distributed Sensor Networks 2015.

[11] K.S. A. Fathima and K. Sindhanaiselvan. "Ant colony optimization-based routing in wireless sensor networks." Int. J. Advanced Networking and Applications 4.04 2013, pp. 1686-1689.

[12] P. Gulia, and Sumita Sihag. "Enhance Security in MANET using Bacterial Foraging Optimization Algorithm." International Journal of Computer Applications 84.1 2013, pp. 32-35.

[13] Ying He,Jian Wang,Liang-Xi Qin;Lin Mei,"A H-K Clustering Algorithm Based On Esemble Learning Smart And Sustainable City ", International Conference (IEEE)(2013), pp.300-305.

[14] S. K. Chaurasiya, T. Pal, and S. D. Bit, "An enhanced energy-efficient protocol with static clustering for wsn," in Information Networking (ICOIN), 2011 International Conference on. IEEE, 2011, pp. 58–63.

[15] D. R. Prasad, P. Naganjaneyulu, and K. S. Prasad, "Modified leach protocols in wireless sensor networksUa review," in ˚ Proceedings of 2nd International Conference on Micro-Electronics, Electromagnetics and Telecommunications. Springer, 2018, pp. 681–688.

[16] Zhang, Qikun, Yongjiao Li, Quanxin Zhang, Junling Yuan, Ruifang Wang, and Yong Gan. "A self-certified cross-cluster asymmetric group key agreement for wireless sensor networks." Chinese Journal of Electronics 28, no. 2 (2019): 280-287.

[17] Zakariayi, Somaieh, and Shahram Babaie. "DEHCIC: A distributed energy-aware hexagon-based clustering algorithm to improve coverage in wireless sensor networks." Peer-to-Peer Networking and Applications 12, no. 4 (2019): 689-704.

[18] Y. Jaradat, M. Masoud, and I. Jannoud, "A mathematical framework of optimal number of clusters in 3d noise-prone wsn environment," IEEE Sensors Journal, pp. 1–1, 2018.

[19] M. A. Al Sibahee, S. Lu, M. Z. Masoud, Z. A. Hussien, M. A. Hussain, and Z. A. Abduljabbar, "Leach-t: Leach clustering protocol based on three layers," in 2016 International Conference on Network and Information Systems for Computers (ICNISC). IEEE, 2016, pp. 36–40.

[20] W. Qian and C. Zanru, "Multi-Hop Routing Based on LEACH in WSN," 2024 IEEE International Conference on Signal Processing, Communications and Computing (ICSPCC), Bali, Indonesia, 2024, pp. 1-6, doi: 10.1109/ICSPCC62635.2024.10770386.

[21] Singh, S., & Sharma, S. (2020). Energy-Efficient Routing Algorithm (EERA) for Prolonging the Lifetime of Wireless Sensor Networks. Journal of Sensor and Actuator Networks, 9(2), 32.

[22] Frank, H., "Shortest paths in probabilistic graphs", Operations Research, August 1969.

[23] Pritsker, A. A. B., "Application of multichannel queueing results to the analysis of conveyor systems", Journal of Industrial Engineering, January 1966.

[24] Mirchandani, P. B., "Shortest distance and reliability of probabilistic networks", Computers & Operations Research, December 1976.

[25] Fishman, G. S., "Estimating critical path and arc probabilities in stochastic activity networks", Naval Research Logistics Quarterly, May 1985.

[26] Adlakha, V. G., "Note—An improved conditional Monte Carlo technique for the stochastic shortest path problem", Management Science, October 1986.

[27] Sigal, C. E., Pritsker, A. A. B., et al., "The stochastic shortest route problem", Operations Research, 1980.

[28] Beigy, H., Meybodi, M. R., "Utilizing distributed learning automata to solve stochastic shortest path problems", International Journal of Uncertainty, Fuzziness and Knowledge-Based Systems, October 2006.

[29] Guo, Y., Li, S., et al., "Learning automata-based algorithms for solving the stochastic shortest path routing problems in 5G wireless communication", Physical Communication, December 2017.

[30] Liu, K., Zhao, Q., "Adaptive shortest-path routing under unknown and stochastically varying link states", IEEE International Symposium on Modeling and Optimization in Mobile, Ad Hoc and Wireless Networks, May 2012.

[31] He, T., Goeckel, D., et al., "Endhost-based shortest path routing in dynamic networks: An online learning approach", IEEE INFOCOM, April 2013.

[32] Chen, W., Wang, Y., et al., "Combinatorial multi-armed bandit: General framework and applications", International Conference on Machine Learning, February 2013.

[33] Russo, D. J., Van Roy, B., et al., "A tutorial on thompson sampling", Foundations and Trends in Machine Learning, 2018.

[34] Wei, Z., Fei, L., et al., "Reinforcement learning for a novel mobile charging strategy in wireless rechargeable sensor networks", International Conference on Wireless Algorithms, Tianjin, China, 2018.

[35] Iima, H., Kuroe, Y., "Swarm reinforcement learning algorithms based on particle swarm optimization", IEEE International Conference on Systems, Man, and Cybernetics, October 2008.

[36] Chen, X., Zhou, H., Li, J., & Wu, Q. (2021). Distributed Load-Balanced Routing in Wireless Sensor Networks. IEEE Transactions on Mobile Computing, 20(5), 1945-1958.

[37] Zhang, Y., & Li, P. (2022). Adaptive Hierarchical Routing for Efficient Communication in Large-Scale Wireless Sensor Networks. Wireless Communications and Mobile Computing, 2022, Article ID 5748306.